\documentclass[aps,
twocolumn,
]{revtex4} 
\usepackage{epsf}
\usepackage{amssymb,amsmath}
\usepackage{hyperref}
\usepackage{graphicx}

\begin{document}
\title{Systematic Microcanonical Analyses of Polymer Adsorption Transitions}
\author{Monika M\"oddel}
\email[E-mail: ]{Monika.Moeddel@itp.uni-leipzig.de}
\author{Wolfhard Janke}
\email[E-mail: ]{Wolfhard.Janke@itp.uni-leipzig.de}
\homepage[\\ Homepage: ]{http://www.physik.uni-leipzig.de/CQT.html}
\affiliation{Institut f\"ur Theoretische Physik,
Universit\"at Leipzig, Postfach 100\,920, D-04009 Leipzig,\\
and Centre for Theoretical Sciences (NTZ), Emil-Fuchs-Stra{\ss}e 1, D-04105 Leipzig, Germany}
\author{Michael Bachmann}
\email[E-mail: ]{bachmann@smsyslab.org}
\homepage[\\ Homepage: ]{http://www.smsyslab.org}
\affiliation{Institut f\"ur Festk\"orperforschung, Theorie II,\\
Forschungszentrum J\"ulich, D-52425 J\"ulich, Germany}%
\begin{abstract}
In detailed microcanonical analyses of densities of states obtained by extensive multicanonical
Monte Carlo computer simulations, we investigate the caloric properties of conformational 
transitions adsorbing polymers experience near attractive substrates. 
For short chains and strong surface attraction, the microcanonical entropy turns out to be a convex 
function of energy in the transition regime, indicating that surface-entropic effects are relevant.
Albeit known to be a continuous transition in the thermodynamic limit of infinitely long chains,
the adsorption transition of nongrafted finite-length polymers thus exhibits a clear signature of a
first-order-like transition, with coexisting
phases of adsorbed and desorbed conformations. Another remarkable
consequence of the convexity of the microcanonical entropy is that the transition
is accompanied by a decrease of the
microcanonical temperature with increasing energy. Since this is a characteristic physical effect
it might not be ignored in
analyses of cooperative macrostate transitions in finite systems. 
\end{abstract}
\maketitle

\section{Introduction}
The advances in processing and manipulating molecules at solid substrates on the nanometer scale
opens up new vistas for technological applications of hybrid organic-inorganic
interfaces. This includes, e.g., the fabrication of nanostructured transistors being sensitive to specific 
biomolecules~\cite{natsume1,lambeth1} and the application of organic electronic devices on polymer basis 
such as organic light-emitting diodes~\cite{geffroy1} and molecular
storage cells~\cite{reed1}\kern0pt. 
Therefore, the investigation of molecular self-assembly~\cite{sarikaya1,gray1} near substrates has
recently been subject of numerous experimental and computational studies, e.g., for peptide adhesion to
metals and semiconductors~\cite{brown1,whaley1,schulten1,goede1,peelle1,goede2,bgbgij1}\kern0pt.
The understanding of the cooperative effects of structure formation at substrates requires 
systematic studies of mesoscopic aspects of adsorption transitions. This regards scaling 
properties near the adsorption/desorption transition in the thermodynamic limit of large 
polymer systems at planar surfaces~\cite{binder1,eisenriegler1,milchev1,metzger1}\kern0pt, 
also under pulling force~\cite{prellberg1,bhattach1}\kern0pt. Adhesion studies of
polymers were also performed at curved surfaces such as
nanotubes~\cite{srebnik1} and nanoparticles~\cite{broo1}\kern0pt. Particular attention
has been dedicated to the complete phase structure of
adsorbed macromolecules which has been investigated by
means of simple lattice models for polymers~\cite{vrbova1,kumar1,bj1,prellberg2,bj2,binder2} and
peptides~\cite{bj3,allen1}\kern0pt, as well as by employing an off-lattice
polymer model~\cite{mbj1}\kern0pt.

Hybrid systems on the nanoscopic scale must basically be considered as being ``small''. Thus, the study
of finite-size effects in the formation of polymer assemblies in a thermal environment is relevant.
Here, we are going to discuss thermodynamic properties of the adsorption transition of a
flexible, interacting polymer at an attractive substrate. 
The polymer is not anchored at the surface and can therefore freely move as long as it does not get into contact with the substrate.
The statistical analysis is performed
in the microcanonical ensemble in order to retain characteristic, non-negligible surface effects. 
This approach has already proven quite useful for a deeper understanding 
of first-order-like structural transitions such as
molecular aggregation processes~\cite{jbj1,jbj2} and protein folding~\cite{chen1,rojas1}\kern0pt. A
particularly striking result was the recent identification
of intrinsic hierarchies of subphase transitions that accompany the overall cooperative process
of assembly~\cite{jbj2}. The relevance of microcanonical thermodynamics~\cite{gross1}
in small-system transitions has also been stated in   
simulational and experimental atomic clustering studies~\cite{labastie1,schmidt1},
fragmenting nuclei~\cite{gross2}\kern0pt,
and for scaling analyses in magnetic systems employing discrete or continuous spin 
models~\cite{gross3,janke1,hueller1,behringer1,behringer2}\kern0pt. A more exotic example is the seminal
application of this approach to astrophysical systems~\cite{thirring1}\kern0pt, which manifests
its broad universality.

The central quantity in the microcanonical formalism is the number (or density) of states $g(E)$
with system energy $E$, or the microcanonical entropy defined as
$S(E)=k_B\ln\, g(E)$, where $k_B$ is the Boltzmann constant. 
In contrast to canonical ($NVT$) statistics, where the temperature $T$ is an externally
fixed control parameter, in the microcanoncial ($NVE$) ensemble it is
\emph{derived} from the entropy, 
$T(E)=[\partial S(E)/\partial E]_{N,V}^{-1}$. In both ensembles, the particle number $N$ and 
the volume $V$ are kept fixed. Particularly interesting microcanonical effects occur in the transition
regime, if the
entropy is a convex function of energy in this region. The physical consequence is that
with \emph{increasing} system energy the temperature \emph{decreases}. This can only be explained
by the fact that conformational transitions of small systems are governed by their surfaces, whereas
volume effects become only relevant for large systems [even a perfectly 
tetrahedral atomic cluster with 309 atoms
contains still more atoms in the outer shell (162) than in the interior (147)]. A remarkable
side effect is the negativity of the microcanonical specific heat, 
$C_V(E)=T(E)[\partial S(E)/\partial T(E)]_{N,V}=[\partial T(E)/\partial E]_{N,V}^{-1}=
-[\partial S(E)/\partial E]_{N,V}^2/[\partial^2 S(E)/\partial E^2]_{N,V}$, in the regions of negative
curvature of $S(E)$.
Thus, as
long as the surface-to-volume ratio is large enough to
suppress a concave increase of the microcanonical entropy
and the energetic separation of the two distinct
phases [one of which is entropy-dominated (e.g., liquid)
and the other energy-dominated (e.g., solid)] is sufficiently large
to establish a kinetic barrier, microcanonical effects matter.
This regards all first-order phase transitions and
two-state systems (e.g., proteins with two-state folding characteristics). 
It also matters for transitions,
where phase co-existence is completely absent
in the thermodynamic limit, but not for the finite systems.
The latter case is what we would like to consider
in more detail in the following: The adsorption transition
of flexible polymers to an attractive substrate, known to
be a second-order phase transition in the thermodynamic
limit. However, as we will show here, the adsorption of nongrafted polymers
with finite lengths exhibits signals of a first-order transition
which we find to vanish in the thermodynamic limit. 
\begin{figure}[t]
\includegraphics[width=7.8cm]{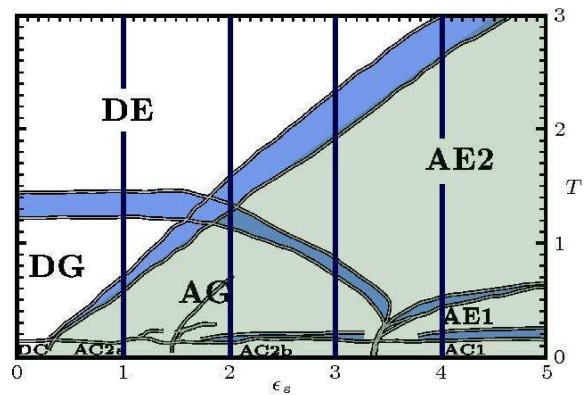}
\caption{\label{fig:pd} Pseudo-phase diagram of a homopolymer with 20
	monomers as obtained in extensive simulations; details are discussed
	in Ref.~\cite{mbj1}. The bands separate the individual
	conformational phases, the band width indicates the statistical
	uncertainty. DE, DG, and DC denote bulk phases of
	expanded coils, globular, and crystalline structures, respectively.
	DE and DG are separated by the $\Theta$-transition. AE1
	is dominated by adsorbed single-layer (two-dimensional) expanded
	structures, AE2 by adsorbed conformations extending
	into the bulk. AG denotes the adsorbed globular regime
	and the crystalline phases differ in their topology (AC1: two-dimensional,
	AC2a,b: three-dimensional). 
	In this work, we will primarily 
	focus on the adsorption
	transition between DE and AE2. Vertical lines are
	placed at values of surface-attraction strengths $\epsilon_s$ chosen for
	the subsequent discussion of microcanonical effects accompanying this transition.}
\end{figure}

\section{Model and Methods}
For our analysis, we consider a single flexible and nongrafted
linear homopolymer with $N$ monomers
that interacts with an attractive planar substrate. In Ref.~\cite{mbj1}, we
have already employed this hybrid model for the strictly canonical identification of conformational 
adsorption phases. Here, we concentrate ourselves on the adsorption/desorption transition between
desorbed (DE) and three-dimensional adsorbed (AE2) expanded conformations (see Fig.~\ref{fig:pd}).  
The polymer is represented by a bead-stick model with standard Lennard-Jones (LJ)
interaction between nonbonded monomers, mimicking short-range volume exclusion and long-range 
van der Waals 
(vdW) 
attraction. Bond lengths between adjacent monomers are normalized to unity.
Since the original model (called the ``AB model''~\cite{stillinger1,baj1}) was designed 
for mesoscopic heteropolymers, an additional weak bending energy was introduced which is kept here. 
The only degrees of freedom in our polymer model are thus the angles between successive bonds. 
Center-of-mass 
translation is restricted to a cavity bounded by the attractive substrate located
at $z=0$ and a sufficiently distant steric wall at $z=L_z$ to prevent 
the polymer from escaping. 
Later in this paper we will discuss the dependence of the microcanonical results on $L_z$ in detail.
If not mentioned otherwise, the monomer density is kept constant, i.e., $L_z$ 
scales linearly with the length of the chain $N$ (we chose $L_z=3N$, i.e., a 
constant concentration of monomers).  
Translation in the $xy$-plane parallel to the walls is irrelevant here.
The interaction
of a monomer with the continuous flat surface of a substrate 
filling the half-space $z\leq 0$ is obtained by integrating a 
12-6-LJ potential over that half-space, which results in 
a 9-3-LJ-like potential~\cite{steele1,mbj1}\kern0pt. 
In our simulations, all lengths are measured in units of the vdW radius
$\sigma=2^{-1/6} r_{\rm min}$, where $r_{\rm min}$ is the minimum of the 12-6-LJ potential,
and energies
in units of a global energy scale $\epsilon_0$. Thus, the temperature scale is given by $\epsilon_0/k_B$. 
Accordingly, for simplicity, we set in the following $\sigma=\epsilon_0=k_B\equiv 1$.
The energy of the hybrid system is then written as:
\begin{eqnarray}
E & = & 4\sum_{i=1}^{N-2}\sum_{j=i+2}^{N}\left( {r_{ij}^{-12}}-{r_{ij}^{-6}}\right)+
\frac{1}{4}\sum_{i=1}^{N-2}\left[ 1-\cos\left(\vartheta_{i} \right)\right]\nonumber\\
& & +\, \epsilon_{s}\sum_{i=1}^{N}\left(\frac{2}{15} {z_i^{-9}} - {z_i^{-3}}\right),
\end{eqnarray}
where $0\leq \vartheta_{i}\leq\pi$ denotes the bending angle between monomers $i$, $i+1$, and $i+2$.
The distance between the monomers $i$ and $j$ is ${r}_{ij}=|\vec{r}_{j}-\vec{r}_{i}|$ and
$z_i$ is the distance of the $i$th monomer from the substrate. The free parameter $\epsilon_s$ 
represents the surface attraction strength and weighs the energy scales 
of monomer-surface ($E_{\rm surf}$) and intrinsic monomer-monomer ($E_{\rm bulk}$) interaction. 

\begin{figure}[t]
\centerline{\includegraphics[width=7.8cm]{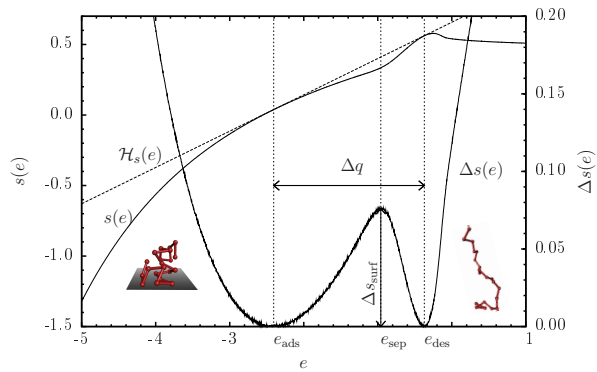}}
\caption{\label{fig:s20} Microcanonical entropy $s(e)$ (up to an unimportant constant)
for a 20mer at $\epsilon_s=5$, the Gibbs hull $\mathcal{H}_s(e)$, and the difference
$\Delta s(e)=\mathcal{H}_s(e)-s(e)$ as functions of the energy per monomer $e$.
The convex adsorption regime is bounded by the energies $e_{\rm ads}=-2.412$ and $e_{\rm des}=-0.369$ of the 
coexisting phases of adsorbed and desorbed conformations at the adsorption temperature $T_{\rm ads}=3.885$, 
as defined via the slope of $\mathcal{H}_s(e)$. The maximum of $\Delta s(e)$, called surface entropy
$\Delta s_{\rm surf}$, is found at $e_{\rm sep}=-0.962$, which defines the energy of phase separation.
The latent heat $\Delta q$ is defined as the energy being necessary to cross the transition region
at the transition temperature $T_{\rm ads}$.}
\end{figure}

Simulations of this model were performed with multicanonical Monte Carlo sampling~\cite{muca1}\kern0pt,
a generalized-ensemble method which directly yields an estimate for the density of states $g(E)$. 
Details of this method, applied to this model, have already been described in Ref.~\cite{mbj1}.
Exemplified for a polymer with 20 monomers and a surface attraction strength $\epsilon_s=5$, we show 
in Fig.~\ref{fig:s20} the microcanonical entropy per monomer
$s(e)= N^{-1} \ln g(e)$ as a function of the energy per monomer $e=E/N$.
It shows the characteristic microcanonical features of a transition with phase coexistence in
a small system. For energies right below $e_{\rm ads}$, the system is in the adsorbed phase 
AE2 (cf.~Fig.~\ref{fig:pd}), i.e., 
the polymer is in contact with the substrate, but monomer-monomer contacts are not particularly
favored and thus expanded conformations dominate. For energies between $e_{\rm ads}$ and
$e_{\rm des}$, the system is in the transition region, where $s(e)$ is convex. This is clearly 
seen by constructing the Gibbs hull 
$\mathcal{H}_s(e)=s(e_{\rm ads})+e(\partial s/ \partial e)_{e=e_{\rm ads}}$ as the tangent that touches 
$s(e_{\rm ads})$ and $s(e_{\rm des})$. Thus,
$T_{\rm ads}=(\partial \mathcal{H}_s/\partial e)^{-1}=(\partial s/ \partial e)^{-1}_{e=e_{\rm ads}}=
(\partial s/\partial e)^{-1}_{e=e_{\rm des}}$ is the microcanonical \emph{definition} of the
adsorption temperature, which
coincides with the temperature determined in canonical
simulations by the often employed criterion of two equal-height peaks in the energy distribution~\cite{janke1}.
However, due to the convex well of $s(e)$, the definition of a single transition
temperature is misleading; the transition rather spans a region of temperatures. Equivalently, for small
systems in the canonical ensemble, fluctuation maxima or two equal-weight peaks
are located at different temperatures~\cite{kotecky1}
which also renders the definition of a unique transition temperature impossible. 
This is obvious
for systems, where the thermodynamic limit is unreachable, such as proteins~\cite{bj4}.
A unique definition of 
the transition point in general is only possible in the thermodynamic limit.

Let us define the deviation between $s(e)$ and $\mathcal{H}_s(e)$ by 
$\Delta s(e)=\mathcal{H}_s(e)-s(e)$. Then, the surface (or interfacial) entropy, 
which represents the entropic barrier of the two-state transition,
is defined as the maximum deviation 
$\Delta s_{\rm surf}=\max\{\Delta s(e)\, |\, e_{\rm ads}\le e\le e_{\rm des}\}$. The peak is located
at $e=e_{\rm sep}$ and defines the energetic phase-separation point.
Finally, the energetic gap between the two macrostates is the latent heat per monomer, 
$\Delta q=e_{\rm des}-e_{\rm ads}=T_{\rm ads}[s(e_{\rm des})-s(e_{\rm ads})]$. 
In the thermodynamic limit, a first-order phase transition
will be characterized as usual by $\lim_{N\to\infty}\Delta q = {\rm const} > 0$, whereas 
$\lim_{N\to\infty}\Delta q = 0$ in the case of a second-order transition. However, in both cases
we expect the surface entropy to vanish in this limit, $\lim_{N\to\infty}\Delta s_{\rm surf} = 0$,
i.e., the microcanonical entropy is always a concave function of energy for infinitely large systems.
Before we show for the adsorption transition that the latent heat indeed decreases with system size,
we first investigate the origin of the phase separation for chains of finite length and 
discuss the adhesion strength dependence of surface entropy and microcanonical temperature.
\begin{figure}[t!]
\centerline{\includegraphics[width=7.8cm]{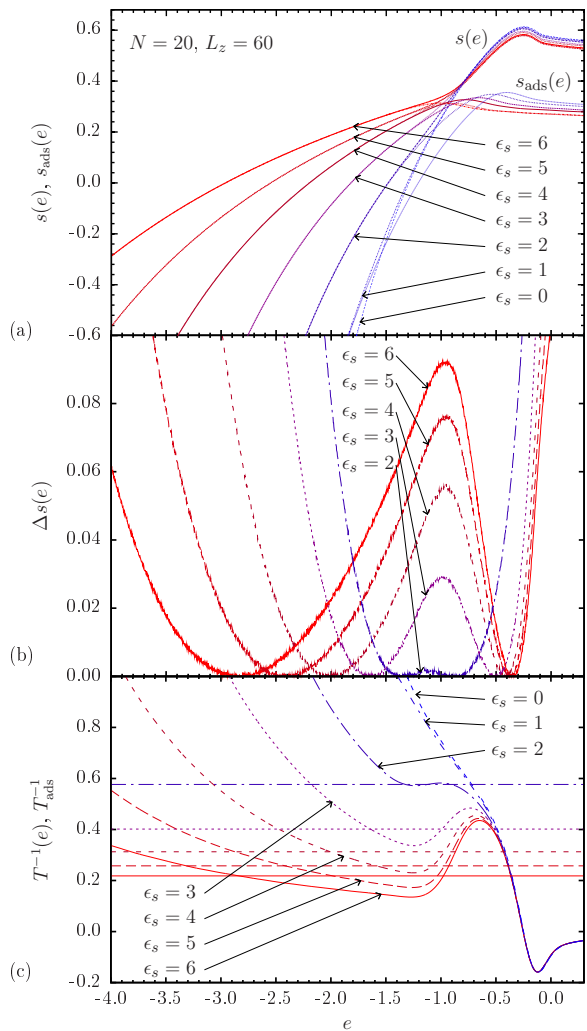}}
\caption{\label{fig:eps} 
(a) Microcanonical entropies and its fraction for adsorbed conformations $s_{\rm ads}(e)$
at various surface attraction strengths $\epsilon_s=0,1,\ldots,6$ for a 20mer 
[the fraction of desorbed structures corresponds to $s(e)$ for $\epsilon_s=0$];
(b) deviations $\Delta {s}(e)$ from the respective Gibbs hulls (not shown) to illustrate the increase
of the surface entropy $\Delta s_{\rm surf}$ and the latent heat $\Delta q$ 
with the attraction strength $\epsilon_s$. Note that
$\Delta s_{\rm surf}=\Delta q=0$ for $\epsilon_s=0,1$; 
(c) caloric inverse temperature curves $T^{-1}(e)$ and Maxwell lines at respective 
reciprocal transition temperatures $T^{-1}_{\rm ads}$.}
\end{figure}

\section{Results}
\subsection{Dependence on the Surface Attraction Strength}
In Fig.~\ref{fig:eps}(a), the microcanonical entropy ${s}(e)$ is shown for a 20mer,
parametrized by the surface attraction strength $\epsilon_s$. Since the high-energy regime is
dominated by desorbed conformations, the density of states and 
hence $s(e)$ are hardly affected by changing the values of $\epsilon_s$. 
The low-energy tail, on the other hand, increases significantly with $\epsilon_s$. 
Thus, it is useful 
to also 
split the density of states into the contributions of 
desorbed and adsorbed conformations, $g_{\rm des}(e)$
and $g_{\rm ads}(e)$, respectively,
such that $g(e)=g_{\rm des}(e)+g_{\rm ads}(e)$ and $s_{\rm des,ads}(e)=N^{-1}\ln g_{\rm des,ads}(e)$.
We define the polymer to be adsorbed if its total surface energy is 
$E_{\rm surf}<-0.1\, \epsilon_s\, N$.
Since $s_{\rm des}(e)$ corresponds to $s_{\epsilon_s=0}(e)$, for $\epsilon_s>0$
only the $s_{\rm ads}(e)$ curves were added in Fig.~\ref{fig:eps}(a). 
Both, $s_{\rm ads}(e)$ and $s_{\rm des}(e)$, are concave in the whole energy range of
the adsorption transition. Thus, convex entropic monotony can only 
occur in the most sensitive region where adsorbed and desorbed conformations
have equal entropic weight, i.e., at the entropic transition point. 
Note that for a polymer \emph{grafted} at the substrate the translational entropy would be very small.
The thus far less pronounced increase of $s_{\rm des}(e)$ at the entropic transition point
is not sufficient to induce the convex intruder and
no microcanonical peculiarities appear in this case. 
 
Depending on $\epsilon_s$ and thus on the energetic location 
of the crossing point, the adsorption transition appears to be \emph{second-order-like} 
($\Delta q=0$ for $\epsilon_s\lessapprox 2$) or \emph{first-order-like} 
($\Delta q>0$ for $\epsilon_s\gtrapprox 2$) 
for a finite, nongrafted chain. 
Referring to the phase diagram
in Fig.~\ref{fig:pd}, the first scenario corresponds to the docking/wetting transition
from desorbed globules (DG) to adsorbed globules (AG). 
The $T^{-1}(e)$ curves for $\epsilon_s=0, 1$ in Fig.~\ref{fig:eps}(c) do not at all exhibit
microcanonical signatures for a first-order-like character of the adsorption transition which occurs
for $\epsilon_s=1$, e.g., near $T_{\rm ads}\approx 0.7$ (see Fig.~\ref{fig:pd}). 
Noticeably, the inflection points near  $T_\Theta^{-1}\approx 0.77$ ($T_\Theta\approx 1.3$)
indicate the $\Theta$-transition that separates coillike and globular conformations
in the bulk (DE/DG). It is a surprising observation that the adsorption
transition becomes first-order-like at the point, where it falls together with the $\Theta$-transition
($\epsilon_s\approx 1.8$, $T\approx 1.3$). This is signaled by the saddle point of the corresponding
$T^{-1}$ curve in Fig.~\ref{fig:eps}(c).

For larger values of $\epsilon_s$, phase coexistence is apparent for the transition 
from expanded coils (DE) to adsorbed coils (AE2). Here,
the corresponding deviations $\Delta s(e)$ from the Gibbs hulls,
as plotted in Fig.~\ref{fig:eps}(b), become maximal at the crossing point. 
The curves of the inverse microcanonical temperatures decrease [i.e., $T^{-1}$ as
plotted in Fig.~\ref{fig:eps}(c) increases] with increasing energy in the transition region. 
The temperature ``bends back'', i.e., in the desorption process
from AE2 to DE, the system is cooled while energy
is increased. This is a characteristic feature of first-order-like 
transition behavior of a finite system (called ``backbending effect''~\cite{gross1})
and has also been observed in numerous other
systems, such as, e.g., peptide aggregation~\cite{jbj1,jbj2}\kern0pt. In Fig.~\ref{fig:eps}(c), 
also the Maxwell lines $T^{-1}_{\rm ads}$ 
[the slopes of the corresponding Gibbs constructions in Fig.~\ref{fig:eps}(a)] are inserted. 
The adsorption temperatures $T_{\rm ads}$ found with this construction depend
roughly linearly on the surface attraction strengths, as it has already been suggested by our formerly
constructed phase diagram in Ref.~\cite{mbj1}. 
Thus, the intersections of the Maxwell lines and the $T^{-1}$ curves are identical with the 
extremal points in Fig.~\ref{fig:eps}(b) and are located at the respective energies
$e_{\rm ads}$, $e_{\rm sep}$, and $e_{\rm des}$. 
It is obvious that the desorption energies
per monomer, $e_{\rm des}$, converge very quickly to a constant value 
$e_{\rm des}^{\rm \epsilon_s\to\infty}\approx -0.35$, when increasing the adhesion strength 
$\epsilon_s$. On the other hand, the adsorption energies $e_{\rm ads}$ still change rapidly.
In consequence, the latent heat per monomer, $\Delta q$, increases similarly
fast with $\epsilon_s$, i.e., the energetic gap between the coexisting macrostates becomes
larger as also does the surface-entropic barrier $\Delta s_{\rm surf}$ [cf.\ Fig.~\ref{fig:eps}(b)].
Since linearly depending on $\epsilon_s$, $e_{\rm ads}$ and $\Delta q$ trivially diverge for
$\epsilon_s\to\infty$. 

\begin{figure}[t!]
\includegraphics[width=7.8cm]{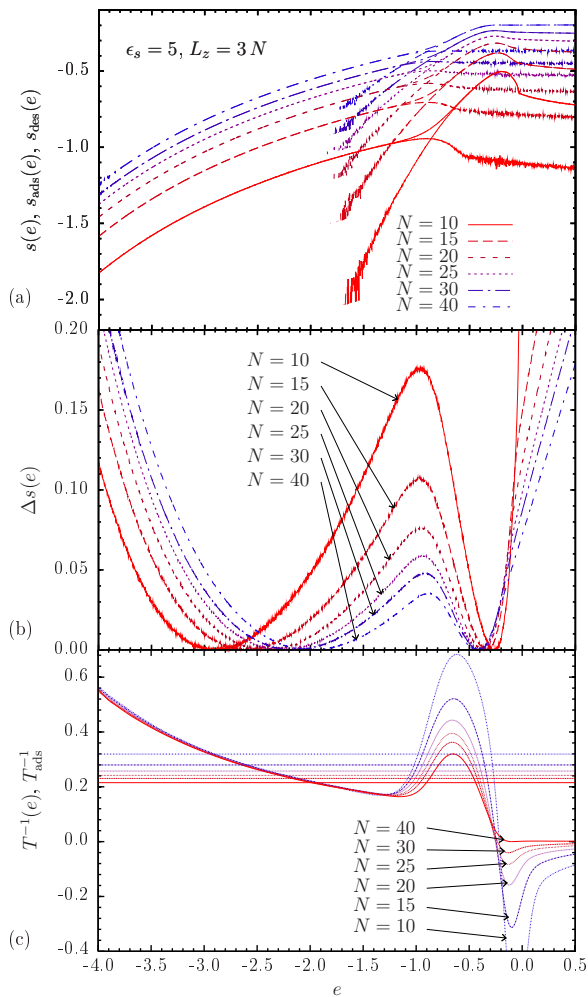}
\caption{\label{fig:N} (a) Microcanonical entropy $s(e)$, adsorption entropy $s_{\rm ads}(e)$, 
and desorption entropy $s_{\rm des}(e)$ 
for polymers with different chain lengths $N=10\ldots 40$ and fixed surface 
attraction strength $\epsilon_s=5$ in the adsorption transition regime. 
The maximum of $s(e)$ and the ``convex intruder'' begin to disappear with increasing chain 
length---a first indication of the tendency of the adsorption transition to
change its characteristics from first-order-like to second-order behavior in the
thermodynamic limit; (b) Deviations $\Delta s(e)$ of $s(e)$ from the Gibbs construction;
(c) caloric inverse temperature curves $T^{-1}(e)$ and Maxwell lines at
$T^{-1}_{\rm ads}$, parametrized by chain length $N$.}
\end{figure}

\subsection{Chain-length Dependence}
Since the adsorption transition between DE and AE2 is expected to be of second order
in the thermodynamic limit\cite{eisenriegler1}, the microcanonical effects and first-order
signatures, as found for the finite system, must disappear for the infinitely large system
$N\to\infty$. Therefore, we now investigate the chain-length dependence of the
microcanonical effects in comparative microcanonical analyses. Fig.~\ref{fig:N}(a) shows the 
microcanonical entropies $s(e)$, the adsorption entropies $s_{\rm ads}(e)$, and the 
desorption entropies $s_{\rm des}(e)$ for chain lengths $N=10,\ldots,40$. The respective
slopes of $s_{\rm ads}(e)$ and $s_{\rm des}(e)$ near the crossing points converge to
each other with increasing chain length. Hence, the depth of the convex well is getting 
smaller and thus also the surface entropy decreases [Fig.~\ref{fig:N}(b)]. Interestingly, 
the separation energies $e_{\rm sep}\approx -0.95$ [which corresponds
to the maxima of $\Delta s$ in Fig.~\ref{fig:N}(b) and 
approximately 
to the location of the
intersection points of $s_{\rm ads}(e)$ and $s_{\rm des}(e)$ in 
Fig.~\ref{fig:N}(a)] do not depend noticeably
on $N$. The desorption energies $e_{\rm des}$ move little, but the 
adsorption energies $e_{\rm ads}$ shift much more rapidly towards the separation point, i.e., 
the latent heat decreases with increasing system size. Consequently, in Fig.~\ref{fig:N}(c),
the backbending of the (reciprocal) caloric temperatures is getting weaker; the adsorption 
temperatures converge towards a constant.
Note that the microcanonical temperature of these finitely long chains is negative in 
the high-energy region. 
This is another characteristic 
feature of finite systems in the microcanonical analysis and disappears with increasing chain lengths, as can be seen
in Fig.~\ref{fig:N}(c).
\begin{figure}[t!]
\centerline{\includegraphics[width=7.85cm]{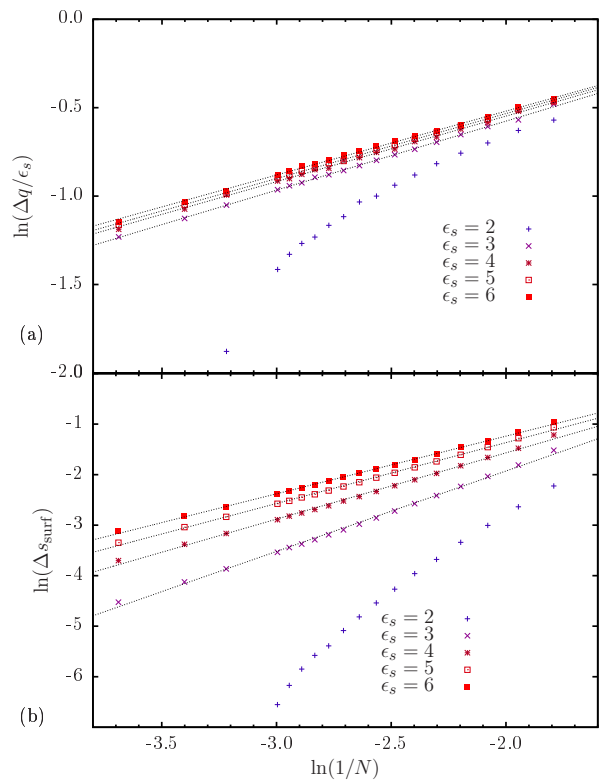}}
\caption{\label{fig:scal} Scaling with polymer length $N$: 
(a) Latent heat per monomer normalized by the surface attraction
$\Delta q/\epsilon_s$ vs.~inverse chain length $1/N$ for several surface attraction strengths $\epsilon_s$  and least-square fit curves
to $\Delta q/\epsilon_s\sim N^{-\kappa_q}$. The data collapse to a single straight line for  
not too small $\epsilon_s$. 
(b) Surface entropy per monomer $\Delta s_{\rm surf}$
vs.~inverse chain length $1/N$
and fits to $\Delta s_{\rm surf}\sim N^{-\kappa_s}$. 
}
\end{figure}

Putting all these informations together, we indeed observe a clear qualitative tendency of the
reduction of the microcanonical effects for larger chains. The rapid decreases
of latent heat and surface entropy qualitatively indicate that 
the adsorption transition of expanded polymers (DE to AE2) crosses over from bimodal first-order-like behavior
towards a second-order phase transition in the thermodynamic limit.

In Fig.~\ref{fig:scal}, the chain-length dependences of the surface entropies $\Delta s_{\rm surf}$ 
and of the latent heats $\Delta q$ are plotted, parametrized by the surface attraction strength 
$\epsilon_s$. The chains considered in our study are too short for a detailed
finite-size scaling analysis. However, for $\epsilon_s>2$, the plots suggest a power-law
dependence of these quantities in this regime. A simple scaling ansatz for the
surface entropy is $\Delta s_{\rm surf}\sim N^{-\kappa_s}$, while for the latent heat
that trivially scales with $\epsilon_s$, we choose $\Delta q/\epsilon_s\sim N^{-\kappa_q}$.
The least-square fits of the data 
yield $\kappa_s=1.65\,\, (1.36,\, 1.24,\, 1.17)$ and $\kappa_q=0.39\,\, (0.37,\, 0.37,\, 0.36)$
for $\epsilon_s=3\,\, (4,\, 5,\, 6)$. The fit curves are also
inserted into Fig.~\ref{fig:scal}. The fit results
for the exponents
depend on $\epsilon_s$, but seem to converge
to constant positive values for $\epsilon_s\to\infty$. 
The surface entropy vanishes in the thermodynamic limit independently of the transition characteristics.
However, that our data suggest $\lim_{N\rightarrow\infty}\Delta q=0$ is support for the assumption of 
the second-order nature of the adsorption transition.
This is 
consistent with results discussed in Ref.~\cite{binder1}.

\subsection{Variation of the Box Size}
\begin{figure}[t!]
\centerline{\includegraphics[width=7.75cm]{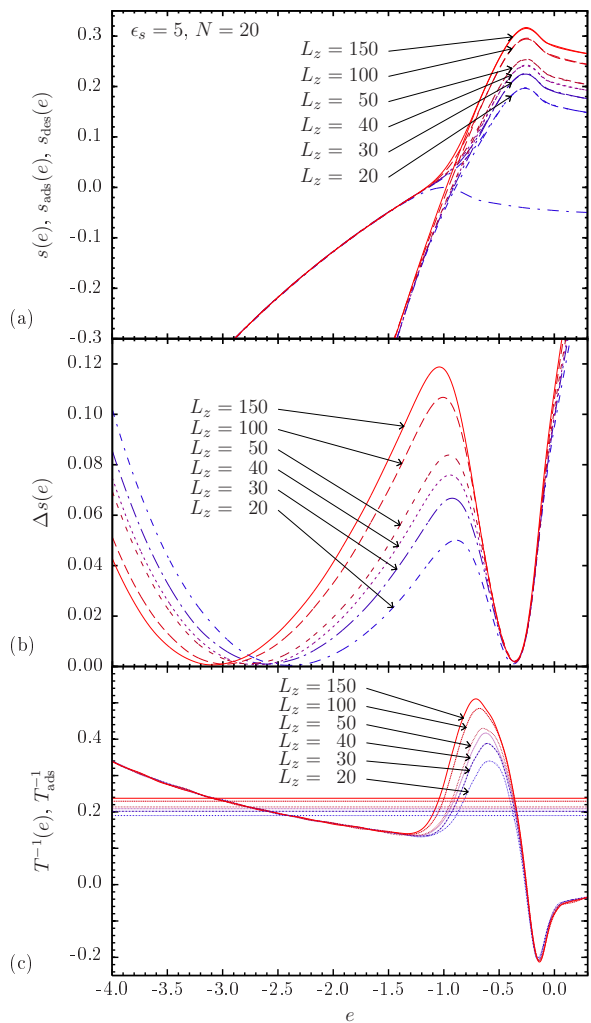}}
\caption{\label{fig:l20} 
(a) Microcanonical entropies and its fractions for adsorbed and desorbed conformations, 
$s_{\rm ads}(e)$ and $s_{\rm des}(e)$, for increasing simulation box size $L_z=20, \ldots, 150$.
The shape of both fractions stays unchanged for different box sizes. Only the amount of desorbed 
conformations increases relative to adsorbed ones for larger boxes;
(b) deviations from the respective Gibbs hulls $\Delta s(e)$. An increased $s_{\rm des}(e)$ induces an increase
of the surface entropy $\Delta s_{\rm surf}$ and slightly also of the latent heat $\Delta q$.
(c) caloric inverse  temperature curves $T^{-1}(e)$ and Maxwell lines, parametrized by the distance
between attractive and steric wall $L_z$.
}
\end{figure}

Finally, after noticing that there is a considerable influence of the simulation box size on the microcanonical
properties of the adsorption transition,
we also want to investigate this effect in more detail. To this end, 
simulations with $\epsilon_s=5$ for a fixed chain length ($N=20$) were performed for different distances $L_z$ of the 
steric wall to the attractive substrate. 
Note that fixing the chain length 
$N$, but changing $L_z$ will also change the density. Hence, the limit of $L_z\rightarrow\infty$
considered in the following does not correspond to the thermodynamic limit.
Analogously to Fig.~\ref{fig:eps} and Fig.~\ref{fig:N}, the corresponding microcanonical results
are displayed in Fig.~\ref{fig:l20}. 
Because the number of adsorbed conformations cannot depend on the amount of space available far away 
from the substrate, the unknown additive constants to $s(e)$, $s_{\rm ads}(e)$, and $s_{\rm des}(e)$ are 
chosen in such a way that $s_{\rm ads}(e)$ coincides for all $L_z$ in Fig.~\ref{fig:l20}(a). It is also possible to overlap
all $s_{\rm des}(e)$ via a suitable additive constant. Hence, the {\em conformational} entropy does not depend
on the simulation box size as long as the simulation box exceeds the chain size. This should not be surprising,
since once all possible conformations can be adopted, there is nothing more to gain.
All what should happen is a gain of {\em translational} entropy proportional to the logarithm of the simulation box size
for desorbed conformations. This is exactly what the data confirm. 
In Fig.~\ref{fig:l20}(b) the consequence of this on $\Delta s(e)$ is shown. Both, the surface entropy $\Delta s_{\rm surf}$
and the latent heat $\Delta q$ increase with $L_z$. 
It is a significant qualitative difference compared to the previous analysis of the limit $N\rightarrow\infty$
that the latent heat remains finite for large box sizes, i.e., $\lim_{L_z\rightarrow\infty}\Delta q\neq 0$.
Thus, the adsorption transition of the finite polymer preserves its first-order-like character in this limit.
The entropic barrier can grow arbitrarily large 
for large simulation boxes since the part of the phase space in proximity of the attractive substrate gets arbitrarily small.
It is interesting to note here that in simulations of the grafted case no intruder was observed.

The resultant caloric inverse temperature curves $T^{-1}(e)$ in Fig.~\ref{fig:l20}(c) only differ in the 
energy regime, where both entropic contributions, $s_{\rm ads}(e)$ and $s_{\rm des}(e)$, are of the same order of
magnitude -- the coexistence region. Once again, the effect of the intruder gets enhanced with $L_z$
and only in this regime $T(e)$ changes with $L_z$. In Fig.~\ref{fig:l20}(c), also the Maxwell lines representing 
the adsorption temperatures are shown.

One can use the knowledge of the behavior of $s_{\rm ads}(e)$ and $s_{\rm des}(e)$ to venture a simple
estimate of $T_{\rm ads}$ by performing a Gibbs construction. 
In the adsorption phase, the contact point of the Gibbs hull is independent of $L_z$, 
$s(e_{\rm ads})=s^{\rm trans,\parallel}(e_{\rm ads})+s^{\rm conf}(e_{\rm ads})$,
where $s^{\rm trans,\parallel}(e_{\rm ads})$ is the translational entropy parallel to the substrate and 
$s^{\rm conf}(e_{\rm ads})$ the conformational entropy of the adsorbed conformations. 
The other contact point $s(e_{\rm des},L_z)=s^{\rm trans,\perp}(e_{\rm des},L_z)+s^{\rm trans,\parallel}(e_{\rm des})+s^{\rm conf}(e_{\rm des})$,
corresponding to the entropy in the desorption phase, is a decomposition of the $L_z$-dependent translational
entropy $s^{\rm trans,\perp}(e_{\rm des},L_z)=N^{-1}\ln L_z$, and the $L_z$-independent contributions from the 
translation parallel to the substrate $s^{\rm trans,\parallel}(e_{\rm des})$ and the conformational entropy $s^{\rm conf}(e_{\rm des})$.
The adsorption temperature is obtained as the inverse slope of the Gibbs hull
\begin{equation}
\label{eq:tadsA}
T_{{\rm ads}}=\frac{\Delta q}{s^{\rm conf}(e_{\rm des})-s^{\rm conf}(e_{\rm ads})+N^{-1} \ln L_z}. 
\end{equation}
The conformational entropies and $\Delta q$ also contain an $N$-dependence, but since we fixed $N$ that
is of no interest here.

For practical purposes, the simple relation~(\ref{eq:tadsA}) 
allows one to restrict oneself to perform a single simulation within a sufficiently large and 
finite box,
and one only has to keep in mind
the simple $\ln(L_z)$ dependence on the simulation box size perpendicular to the substrate.

\section{Summary}
In our study, we have investigated the adsorption transition of polymers at attractive
substrates with different binding strengths by means of microcanonical analyses. We
found that the adsorption transition 
exhibits clear signals of a first-order-like conformational transition in the important case of finitely long
polymers, whereas it crosses over into a second-order phase transition in the thermodynamic
limit, as expected. As the microcanonical analysis revealed furthermore, the first-order-like
character of the finite-size effects is expressed by the coexistence of adsorbed and desorbed
conformations at the adsorption transition temperature which is defined by the Maxwell
construction in the transition region of the caloric temperature curves. The energetic separation of these 
phases is defined by the intersection of the caloric temperature curve and the Maxwell
line in the region, where temperature decreases with increasing energy. This typical
microcanonical effect in a finite system is due to dominant effects at the polymer coil surface.
Consequently, surface entropy and latent heat
are nonzero for finite systems, but vanish in the thermodynamic limit. We performed scaling
analyses for the decrease of these quantities and found that surface entropy and latent heat
decay slower for larger surface attraction strength. 
Also the simulation box dependence of the intruder and the adsorption temperature was analyzed.
We found that it is quite substantial, but of a rather trivial nature that only affects the desorbed
chains via an additional translational entropy proportional to the logarithm of the box size.
All in all, our study has shown the usefulness
of the microcanonical interpretation in the particularly interesting case of
the adsorption transition of a nongrafted polymer.

\section{Acknowledgements}
This work is partially supported by the DFG (German Science 
Foundation) within the Graduate School BuildMoNa and 
under Grant Nos.\ JA \mbox{483/24-1/2/3}, the Deutsch-Franz\"osische Hochschule 
(DFH-UFA) under Grant No.~CDFA-02-07, and by the German-Israel Program ``Umbrella''
under Grant No.\ SIM6. 
Support by supercomputer time grants (Grant Nos.~hlz11 and JIFF39) of the 
Forschungszentrum J{\"u}lich is gratefully acknowledged.
\vspace*{0.5cm}

%
\end{document}